\documentclass[12pt]{article}
\usepackage{a4wide}
\usepackage{amssymb}
\begin{document}
{\renewcommand{\thefootnote}{\fnsymbol{footnote}}
\hfill  IGC--08/11--5\\
\medskip
\begin{center}
{\LARGE  Consistent Loop Quantum Cosmology}\\
\vspace{1.5em}
Martin Bojowald\footnote{e-mail address: {\tt bojowald@gravity.psu.edu}}
% and ... 
\\
\vspace{0.5em}
Institute for Gravitation and the Cosmos,\\
The Pennsylvania State
University,\\
104 Davey Lab, University Park, PA 16802, USA\\
\vspace{1.5em}
\end{center}
}

\setcounter{footnote}{0}

\newcommand{\lP}{\ell_{\mathrm P}}
\newcommand{\md}{{\mathrm{d}}}
\newcommand*{\R}{{\mathbb R}}
\newcommand*{\N}{{\mathbb N}}
\newcommand*{\Z}{{\mathbb Z}}
\newcommand*{\Q}{{\mathbb Q}}
\newcommand*{\C}{{\mathbb C}}

\begin{abstract}
 A consistent combination of quantum geometry effects rules out a
 large class of models of loop quantum cosmology and their critical
 densities as they have been used in the recent literature. In
 particular, the critical density at which an isotropic universe
 filled with a free, massless scalar field would bounce must be well
 below the Planck density. In the presence of anisotropy, no model of
 the Schwarzschild black hole interior analyzed so far is consistent.
\end{abstract}

Aside from detailed technical constructions, one of the main
achievements of loop quantum gravity \cite{Rov,ALRev,ThomasRev} is to
provide a framework for discrete dynamical geometries. Loop quantum
cosmology \cite{LivRev} takes these ingredients and applies them to
expanding universes or black hole models. In this way, effects of
quantum physics as well as quantum geometry can be explored in
detail. This is a specific realization of a general issue which has
been discussed recurrently in the context of discrete models of
gravity \cite{Weiss,UnruhTime,River,EvolvingHilbert,CosConst}. Loop
quantum gravity makes many of the older considerations more specific,
but since neither its complete form nor the precise transition to loop
quantum cosmology has been hammered out as of now, the dynamics of
loop quantum cosmology cannot be unique. Current ignorance must be
parameterized so that at least qualitative implications can be
found. In particular, {\em all corrections from quantum physics and
quantum geometry} must be analyzed before reliable and robust
conclusions can be drawn. Here we consider quantum geometry in regimes
where its implications are dominant over genuine quantum corrections,
and provide a consistent combination of its two main effects: holonomy
corrections but also inverse volume corrections which have often been
ignored. As we will see, this combination leads to tight consistency
conditions which rule out parameter choices made so far in cases where
only holonomy corrections were considered. Quantum back-reaction
effects, which still remain to be fully derived following the methods
of \cite{EffAc,EffCons}, will not be required for our analysis.

We start with a discussion of isotropic models, which classically have
only one invariant scale: the Hubble distance ${\cal
H}^{-1}=a/\dot{a}$ (a dot meaning a derivative by proper time). If we
consider a fixed region ${\cal V}$ in our isotropic space, a second
scale $a^3V_0$ arises which is independent of coordinates but depends
on the coordinate size $V_0$ of the region chosen. These two scales
provide the classical canonical pair, $\{{\cal H},V_0a^3\}=4\pi G$
with the gravitational constant $G$.

Quantum geometry effects are expected to arise if the Hubble scale is
Planckian, ${\cal H}^{-1}\sim \lP$. Technically, this is realized in
loop quantum gravity because holonomies, which are nonlinear functions
of $\dot{a}$ and $a$, replace the classical Hubble parameter in the
gravitational Hamiltonian \cite{LoopRep}. But also the intrinsic
spatial geometry, which depends on $a$ but not on $\dot{a}$, implies
corrections when all of space is made of discrete patches whose size
is near Planckian. (By ``patch'' we will mean the smallest building
block of a discrete geometry.) This arises technically because inverse
patch sizes appear in Hamiltonians, and their quantizations differ
from classical values for small enough patches
\cite{QSDI,QSDV}. However, in contrast to the invariant scale
$\dot{a}/a$, there is no invariant measure for the spatial geometry in
a purely isotropic setting: the scale factor $a$ is coordinate
dependent while the coordinate independent size $V_0a^3$ depends on
the region ${\cal V}$ chosen. This is the reason why inverse volume
corrections have not been fully realized in minisuperspace models so
far, and most often were ignored.

The patch size refers to an underlying discrete state giving rise to
the expanding universe. To include this effect in a pure
minisuperspace model, extra input is thus required from the full
theory, which must guide the construction of homogeneous
models. Fortunately, this is possible in a heuristic mean-field
picture \cite{InhomLattice} where a mean-field describing the
underlying discrete structure is used. As always with mean-field
approximations, the precise way in which the field arises from a
fundamental theory may be very difficult to derive. In lieu of a
detailed derivation, however, there are often consistency conditions
which can help to reduce the a-priori freedom involved in the choice
of mean-field behaviors. This turns out to be the case for loop
quantum cosmology, too.

We then have a patchwork structure describing the isotropic
classical geometry in an atomic manner. Instead of the classical size
$a^3V_0$, we can write ${\cal N}v$ where $v$ is the mean size of a
patch and ${\cal N}$ is their number in the region ${\cal V}$. Since
both products are supposed to give the size of the same region, they
must equal each other:
\begin{equation} \label{Nva}
 {\cal N}(t)v(t)=a(t)^3V_0\,. 
\end{equation}
In this equation, we have already accounted for the main result from
loop quantum gravity employed here: the fact that any dynamics
proposed for the full theory \cite{RS:Ham,QSDI,AQGI} must change the
values of discrete contributions to volume; $v(t)$ must depend on time
as nodes representing the atomic volume elements are being excited. In
general, also the number of patches ${\cal N}(t)$ is a function of
time, and the product of both functions provides the time dependence
of the scale factor $a(t)$. (For the current purposes it is irrelevant
what time variable is used. In explicit constructions, this
could be an internal time such as a matter field.)

As an important consequence, we see that the expansion of an isotropic
universe in loop quantum cosmology is described by {\em two} time-dependent
functions, ${\cal N}(t)$ and $v(t)$, rather than one classical
function $a(t)$. Thus, there is more freedom in the underlying
dynamics which is related to the refinement behavior of an underlying
discrete state. By a pure minisuperspace quantization, which starts
from the classical $a$ and turns it into an operator, this freedom
cannot be constrained but rather emerges in the form of quantization
ambiguities in the reduced Hamiltonian. The detailed form of the
dynamical change of ${\cal N}$ and $v$ can in principle be derived
from the full theory, or be restricted by analyzing the implied
phenomenology of models. (It is not uncommon that additional
parameters and functions not seen classically arise from a quantum or 
microscopic treatment. In cosmological models, for instance, a similar
behavior has been observed in \cite{CosmoWithoutInfl}.)

Quantum geometry is discrete but local, and thus corrections to the
classical geometry and its dynamics depend only on the microscopic
patch size $v$ rather than the macroscopic number ${\cal N}$ of
patches. (From a classical perspective, this behavior may not seem
local if a discrete patch such as the geometry given by a single spin
network vertex is viewed as a representation of the continuum
distribution in a neighborhood. But quantum geometry is local in the
sense that geometrical operators only refer to sums over single patch
contributions, a behavior which we will call ``patch-local'' in what
follows.) This already implies that all consistent implementations of
quantum corrections must be independent of the size $V_0$ of the
region, as this is a parameter which determines ${\cal N}$ but does
not affect $v$.

We first address holonomy corrections which are easier to construct,
although also here extra input is needed compared to the original and
direct quantization used in loop quantum cosmology
\cite{IsoCosmo}. Holonomies are non-linear functions of a connection
which in isotropic models is simply proportional to $\dot{a}$:
$\tilde{c}=\gamma \dot{a}$ with the Barbero--Immirzi parameter
$\gamma$ \cite{AshVarReell,Immirzi}. In a patch-local holonomy, the
connection appears in a line integral along a curve of the (linear)
patch size $\ell_0$. In an isotropic setting, holonomies are thus
functions of $\ell_0\tilde{c}=\gamma L \dot{a}/a$ where $L=\ell_0a$ is
the geometrical and coordinate independent length of the curve (or its
co-moving size).

The number of patches of coordinate size $\ell_0^3$ in a region of
size $V_0$ is given by ${\cal N}=V_0/\ell_0^3$. Together with
(\ref{Nva}), this implies that $L=\ell_0a= (V_0/{\cal
N})^{1/3}a=v^{1/3}$ is directly related to the microscopic patch size
$v$. Thus, holonomy corrections, which arise from deviations of
non-linear functions such as $\sin(\gamma L\dot{a}/a)$ from the linear
$\gamma L\dot{a}/a$, are independent of the size $V_0$, as
required. Below we will see that this is also true for inverse volume
corrections.

Before discussing inverse triad corrections and their mutual
consistency with holonomy corrections, we emphasize the dynamical role
of $L(t)=v(t)^{1/3}$. In the original formulation of loop quantum
cosmology \cite{IsoCosmo,Bohr}, it was assumed that $\ell_0$ is
constant as the universe expands. This is the most straightforward
assumption in a pure minisuperspace construction, but it was clear
that this cannot describe long phases of expansion: a constant
$\ell_0=(V_0/{\cal N})^{1/3}$ implies that the number of patches in
the co-moving region ${\cal V}$ is constant. As the geometrical size
of this region is expanded by the scale factor, a constant number of
patches means that patch sizes are expanded to macroscopic
values. This would clearly be in conflict with a large universe free
of noticeable quantum geometry effects. (The correspondence between
constant $\ell_0$ and a constant number of patches is also
demonstrated by the two-patch model of \cite{LQCStepping}.)

To address this problem, an improvised version of the dynamics was
proposed in \cite{APSII}, based on the ad-hoc assumption that
low-lying eigenvalues of the full area spectrum should determine
quantization parameters of minisuperspace models. This assumption is
ad-hoc because the area operator plays no role in full constructions
of Hamiltonian constraints or their solutions. One could think that it
somehow arises in a gauge-fixing procedure which leads one from the
full theory to an isotropic model, but no such construction is
known. (The use of the area operator, which is not a physical
observable, was criticized in \cite{BounceArea}. However, the area is
used when constructing the constraint and before solving it;
non-observables can certainly appear in this step just as $\dot{a}/a$,
which is not a constant of motion, appears in the classical
constraint.) Despite of these shortcomings, the procedure implied that
it is not $\ell_0$ which is constant but $L=\ell_0a=v^{1/3}$. This
means that the microscopic patch size is constant in this version, and
correspondingly the number of patches increases in an expanding
universe. In this way, strong quantum geometry effects in a large
universe can be avoided for suitable parameter choices. However, the
values proposed in \cite{APSII}, motivated by the smallest non-zero
area eigenvalues, turn out to be in conflict with inverse volume
corrections, to which we turn now.

Inverse volume corrections arise because the patch volume $v$, when
quantized, becomes an operator with discrete spectrum containing zero
\cite{AreaVol,Vol2}. Such an operator does not have a densely defined
inverse, but an inverse patch volume is needed in most
Hamiltonians. One can construct operators with the correct classical
limit given by the inverse based on identities such as
\begin{equation}\label{Poisson}
 ie^{i\ell_0c/V_0^{1/3}} \{ e^{-i\ell_0c/V_0^{1/3}} ,|p|^{3r/2}\}=
 4\pi \gamma G \frac{\ell_0}{V_0^{1/3}} |p|^{3r/2-1} {\rm sgn}(p)
\end{equation}
for the canonical pair $c=V_0^{1/3}\tilde{c}=V_0^{1/3}\gamma\dot{a}$ and
$|p|=V_0^{2/3}a^2$, $\{c,p\}=8\pi\gamma G/3$, underlying isotropic
loop quantum cosmology. For $0<r<2/3$, the right hand side gives an
inverse power of the scale factor, while no inverse is needed on the
left. This gives rise to well-defined quantizations, which are part of
a general procedure \cite{QSDV}, and whose specific constructions in
isotropic models all follow the derivations in \cite{InvScale} to
which we refer for further details.

Since the quantized holonomy $\exp(i\ell_0c/V_0^{1/3})$ becomes a
shift operator in $p$ and the Poisson bracket in (\ref{Poisson})
becomes a commutator, the quantization of inverse volume provides a
discrete approximation of the derivative involved in the classical
Poisson bracket. The derivative $\md |p|^{3r/2}/\md p$ on the right
hand side of (\ref{Poisson}) is then replaced by a difference
$(|p+\Delta p|^{3r/2}-|p-\Delta p|^{3r/2})/2\Delta p$ where $\Delta p$
depends on the precise implementation of the quantization scheme and
is typically related to the Planck length. (Numerical studies of the
volume spectrum in the full theory, whose results will shed light on
which precise values are to be expected, are on-going
\cite{VolNum,BoundFull,VolSpecI,VolSpecII}.)  Deviations of the
difference from the derivative give rise to inverse volume
corrections.

The form of $\Delta p$ is important to demonstrate the consistency of
inverse volume corrections, i.e.\ independence of $V_0$, and to
estimate their size. Inverse volume corrections as they have been used
in most isotropic models so far correspond to using the total volume
$V=a^3V_0$ of the region ${\cal V}$ and approximating its derivative
by differences of the form $V\pm \lP^3$. This has two consequences:
(i) inverse volume corrections of this form are extremely tiny when the
total volume is much larger than the Planck volume, and (ii) such
corrections depend on the region ${\cal V}$ and its coordinate size
$V_0$ because $V$ does. The first property led \cite{APSII} to
conclude that inverse volume corrections can safely be ignored
compared to holonomy corrections, but the $V_0$-dependence remained
unexplained; in fact, it points to an inconsistency in the derivations
used which cannot simply be eliminated by ignoring inverse volume
corrections.

In the present context one can see a third problem with such a
treatment of inverse volume corrections: if quantizations of the
inverse volume depend on the total volume of a region, resulting
operators cannot be patch-local. Inverse volume operators appearing
e.g.\ in energy densities would depend on the full volume of a region
${\cal V}$ which can be macroscopic. Although this may not appear as a
problem in an isotropic context where homogeneity identifies local and
global properties, it is clear that such quantizations cannot arise in
a model consistently related to what we know from the full theory. As
already mentioned, full Hamiltonians have contributions which are
patch-local and depend on the patch-volume $v$ but not on the number
of patches ${\cal N}$ or the total volume $V$. This has two important
consequences: (i) consistent implementations of inverse volume
corrections {\em do not depend on $V_0$} since $v$ does not, and (ii)
they are much larger than naively expected because $v$ is much smaller
than $V$ and differences $v\pm\lP^3$ deviate much more from
infinitesimal displacements of $v$ than $V\pm\lP^3$ does from
displacements $\md V$ of the large $V$. That inverse volume
corrections are artificially suppressed in direct minisuperspace
quantizations has also been discussed in the appendix of
\cite{SchwarzN}. 

Consistent inverse volume corrections imply correction functions
$\alpha(L/\lP)$ at all places where inverse densitized triad
components are used in a classical Hamiltonian, such as in the kinetic
term of a matter Hamiltonian \cite{QSDV}. The precise functional form
of $\alpha$ is subject to quantization ambiguities
\cite{Ambig,ICGC,QuantCorrPert}, but is restricted by anomaly
cancellation conditions \cite{ConstraintAlgebra}. In all cases,
$\alpha$ deviates strongly from the classical value $\alpha=1$ when
its argument is of the order one or smaller; for large values of
$L/\lP$ the classical limit is approached asymptotically. Importantly,
correction functions $\alpha(L/\lP)$ are to be evaluated at $L/\lP$
rather than the much larger and $V_0$-dependent $V^{1/3}/\lP$.  There
is thus no $V_0$-dependence, and quantum geometry effects from inverse
volume operators are much more pronounced than expected. In this way,
valuable consistency conditions can arise. In particular, for holonomy
corrections to be small in classical regimes we require
$\ell_0\tilde{c}=\gamma L\dot{a}/a\ll1$, while small inverse volume
corrections require $v^{1/3}/\lP=L/\lP\gg1$. For a given classical
geometry, we thus have an upper as well as a lower bound on $L$, which
leaves only a finite range of allowed choices.

To make this quantitative, we derive the main effect of holonomy
corrections in the current setting, and compare parameter values with
what is consistent with inverse volume corrections. The Hamiltonian
constraint is the main place where corrections enter, and holonomy
corrections imply a form
\begin{equation} \label{Constr}
 C=-\frac{3}{8\pi G\gamma^2} \frac{V_0^{2/3}}{\ell_0^2}\sqrt{|p|}
  \sin^2(\ell_0c/V_0^{1/3})+|p|^{3/2} \rho=0
\end{equation}
with a matter density $\rho$. These terms strictly refer to operators
as they appear in a quantization of the Hamiltonian constraint, but
they can also be considered as the tree-level approximation to the
quantum constraint operator. Our considerations will not require deep
quantum regimes as they are, e.g., encountered around classical
singularities, where such tree-level equations would be unreliable due
to the presence of additional quantum corrections
\cite{BouncePot,BounceSqueezed}. The tree-level constraint then
implies equations of motion with tree-level corrections from quantum
geometry. We will only require the equation
\[
 \dot{p}=\{p,C\}= 2\sqrt{p} \frac{V_0^{1/3}}{\gamma\ell_0}
 \sin(\ell_0c/V_0^{1/3})\cos(\ell_0c/V_0^{1/3})
\]
which can be written as
\begin{eqnarray}
 \left(\frac{\dot{a}}{a}\right)^2 &=& \left(\frac{\dot{p}}{2p}\right)^2
 =\frac{1}{\gamma^2\ell_0^2a^2} \sin^2(\ell_0c/V_0^{1/3})
\left(1-\sin^2(\ell_0c/V_0^{1/3})\right)\\
&=& \frac{8\pi G}{3} \rho\left(1-\frac{8\pi G}{3} \gamma^2\ell_0^2a^2\rho\right)
\end{eqnarray}
where we used the Hamiltonian constraint (\ref{Constr}) to express
$\sin^2(\ell_0c/V_0^{1/3})$ in terms of $\rho$. Defining the critical
density
\begin{equation} \label{rhocrit}
 \rho_{\rm crit}= \frac{3}{8\pi G\gamma^2L^2}
\end{equation}
we thus have the tree-level Friedmann equation \cite{RSLoopDual}
\begin{equation} \label{Tree}
 \left(\frac{\dot{a}}{a}\right)^2 =\frac{8\pi G}{3}
 \rho\left(1-\frac{\rho}{\rho_{\rm crit}}\right)
\end{equation}
whose behavior is determined by the patch size $L$ (see also
\cite{QuantumBounce} for the derivation with general $L$). The meaning
of the critical density $\rho_{\rm crit}$ can be seen in a model where
the only matter ingredient is a free, massless scalar: In this case
the tree-level equation is an exact effective equation
\cite{BouncePert} and can be used even in deep quantum regimes. When
$\rho=\rho_{\rm crit}$, the scale factor reaches a minimum and the
universe bounces. For other matter there are additional quantum
corrections \cite{BounceSqueezed}, and even the tree-level form can
change if higher order holonomy corrections are included which are
possible due to quantization ambiguities \cite{AmbigBounce}. The
bounce has reliably been demonstrated only for a free massless
scalar, but the critical density can nonetheless be used more
generally as a measure for the size of holonomy corrections.

It is hard to derive $L$ from an inhomogeneous state in full loop
quantum gravity, but one can easily check the mutual consistency of
any value for $\rho_{\rm crit}$ in holonomy corrections with the
corresponding implications of $L$ in inverse volume corrections. In
the case of \cite{APSII}, which is the main specific model analyzed
recently, we have a constant $L$ which, by reference to the full area
spectrum, gives rise to a critical density near the Planck density:
$\rho_{\rm crit}\sim\rho_{\rm P}=1/G\lP^2$. With (\ref{rhocrit}) we
obtain $\gamma L\sim \sqrt{3/8\pi} \lP$ and thus $L/\lP\sim
\sqrt{3/8\pi}\gamma^{-1}$. For this value we have $L/\lP>1$, thanks to
the smallness of $\gamma\sim 0.24$ as it arises from black hole
entropy calculations \cite{Gamma,Gamma2}. This is sufficient to avoid
extremely strong inverse volume corrections. However, $L/\lP\sim 1.4$
is not much larger than one and inverse volume corrections for this
value are still significant. In the specific case of constant $L$ as
in \cite{APSII}, the consequences may not be too severe because a
constant $L$ implies that inverse volume corrections merely amount to
a constant factor in all terms of the constraint where an inverse
volume appears. This changes some coefficients, but would leave the
qualitative behavior untouched. However, such a parameter choice would
be unstable: a slight deviation from a constant patch size $L$, as it
can easily occur in general states, would mean that inverse volume
corrections become $a$-dependent. Now, it would not only be
coefficients but even the qualitative dynamical behavior which is
subject to strong changes since, for instance, corrected energy
densities $\rho$ containing $\alpha(L/\lP)$ would depend differently
on the scale factor than they do classically.

In a strict sense with the usual value of the critical density, we
have disproved the improved dynamics of \cite{APSII}.  This dynamics
can be consistent with weak inverse volume corrections only if the
critical density is much lower than Planckian, and the patch size much
larger. For $L$ larger than $\lP$ by one order of magnitude, for
instance, the critical density could at most be about $2\%$ of the
Planck density.

A constant patch size $v$ or a constant $L$ is only a special case,
for which the number of patches is proportional to the volume, ${\cal
N}\propto a^3$. In general, however, both $v$ and ${\cal N}$ can be
functions of $a$ if the scale factor is used as internal time to
measure the change of discrete geometry. As usually, it is often
convenient to assume power-law behaviors at least for certain phases
of the universe evolution, which we parameterize by a power $x$ in
${\cal N}={\cal N}_0a^{-6x}$. The patch size then behaves as
\begin{equation}
 v=a^3V_0/{\cal N}= a^{3(1+2x)}V_0/{\cal N}_0\,.
\end{equation}
Notice that the parameter ${\cal N}_0$ in general depends on the size
of the region ${\cal V}$ as well as coordinates in order to make
${\cal N}$ coordinate independent and proportional to $V_0$; it is not
simply a numerical constant.

In this parameterization, we have $x<0$ if the
number of patches ${\cal N}$ is increasing, and $x>-1/2$ if the patch
size $v$ is increasing. Since there is a lower limit for both ${\cal
N}$ and $v$, values $x>0$ or $x<-1/2$ cannot be realized
forever. Thus, if a uniform power-law model is used, as it is for
instance done in solvable models
\cite{BouncePert,BounceCohStates,BeforeBB,Harmonic} for a free
massless scalar, one has to restrict the power to the generic range
$-1/2<x<0$. The limiting values are possible, but would not be generic
as they would require either no change in the patch size to occur
($x=-1/2$) or a constant number of patches ($x=0$). The former choice,
which gives a constant $L$ and thus the dynamics of \cite{APSII}, is
impossible in full constructions of loop quantum gravity while the
latter is not consistent with large-scale semiclassicality. In
general, power laws can only be assumed for certain periods of time,
as it is commonly done for energy densities in different phases of the
universe, and occasionally a power $x$ may even fall outside the range
$-1/2<x<0$. Different power-laws can then follow each other in much
the same way as, e.g., a dust-dominated phase with one power-law for
energy density would follow a radiation-dominated phase with a
different power-law. Complete functions ${\cal N}(a)$ and $v(a)$
incorporating these different phases could only be derived from a full
solution of an inhomogeneous state which is difficult, but
phenomenological constructions are already possible. See e.g.\
\cite{RefinementInflation,RefinementMatter,tensor,FermionBBN} for
early-universe restrictions on $x$ and $L$. As control over the theory
increases, one can expect tight consistency conditions to arise in
this way.

To restrict general choices even more strongly, anisotropic models are
valuable. Effects of lattice refinements, as first analyzed in this
context in \cite{SchwarzN}, now allow more freedom because parameters
in general depend not only on the volume but also on extensions in
different directions. The interior of a Schwarzschild black hole, for
instance, is of a Kantowski--Sachs geometry with two independent
scales to determine the spatial geometry:
\begin{equation} \label{KS}
 \md s^2= \frac{(p_1)^2}{|p_2|} \md x^2+ |p_2| \md\Omega^2\,.
\end{equation}
There are two densitized triad components $p_1$ and $p_2$, which
provide the volume $V=4\pi L_0 |p_1|\sqrt{|p_2|}$ where $L_0$ is a
coordinate length parameter to select a finite region whose angular
extension has been fixed by using the full sphere size $4\pi$; for
details see \cite{BHInt} or \cite{ModestoConn}. When triad components
and thus the volume vanish, we have either the black hole singularity
($p_2=0$) or the horizon ($p_1=0$).

In this context, we can write a more general equation to relate
microscopic quantities to the classical geometry:
\begin{equation} \label{NLNA}
 {\cal N}_1(t)L(t){\cal N}_2(t)A(t)= 4\pi L_0 |p_1(t)|\sqrt{|p_2(t)|}
\end{equation}
where ${\cal N}_1$ is the number of patches in the $x$-direction, with
patch extension $L$ in this direction, and ${\cal N}_2$ is the number
of patches in the spheres spanned by the angular directions, with
patch area $A$ in these directions. (Spherical symmetry implies that
we do not have to distinguish the two linear extensions in angular
directions, which we thus combine in the angular area.) 

The volume relation (\ref{NLNA}) then factorizes in two identities
\begin{equation}
 {\cal N}_1(t)L(t)= L_0 \frac{|p_1(t)|}{\sqrt{|p_2(t)|}} \quad,\quad
 {\cal N}_2(t)A(t)= 4\pi |p_2(t)|
\end{equation}
using the sizes following from the spatial Kantowski--Sachs metric
(\ref{KS}) in triad variables.  To specify the refinement fully, we
have to determine two functions such as ${\cal N}_1(p_1,p_2)$ and
${\cal N}_2(p_1,p_2)$; just the volume dependence of the total number
${\cal N}_1{\cal N}_2$ of patches is not sufficient. A particular
case, which provides constant patch sizes and shapes, is obtained when
$L$ and $A$ are constant and thus ${\cal N}_1 \propto
|p_1|/\sqrt{|p_2|}$ and ${\cal N}_2\propto |p_2|$. This case was
introduced in \cite{SchwarzN} and shown to have stable and near
classical behavior in a large region of the phase space. It is a
special case of models where the total patch number is proportional to
volume: ${\cal N}_1{\cal N}_2\propto |p_1|\sqrt{|p_2|}\propto V$.

More generally, one can assume a volume dependence ${\cal N}_1{\cal
N}_2\propto V^{-2x}$ which still leaves different choices for the
anisotropic behavior of the patches. What is important in the present
context is that none of these power-law cases can be fully consistent
in classical regimes: The total volume vanishes not only at the
singularity, where strong discreteness effects would not be
unexpected, but also at the horizon where $p_1=0$. For a consistent
model, which behaves semiclassically near the horizon as it should be
the case for massive black holes, one has to use more general
functions than power laws to parameterize the refinement behavior. In
particular, ${\cal N}_1{\cal N}_2$ must not vanish when
$p_1=0$. Despite many attempts, no fully consistent effective
Schwarzschild interior space-time has been constructed yet, even if
one restricts oneself to regions far from classical singularities
where tree-level equations are reliable and quantum back-reaction
effects can be ignored. Anisotropic models thus provide important test
cases which are restrictive enough to constrain lattice refinement
models further.

To conclude, we have seen that a consistent treatment of loop quantum
cosmology, with a robust connection to constructions in the full
theory, requires input from the behavior of an underlying microscopic
state. If this is done correctly, the balance of quantum corrections
changes such that several choices made so far can be ruled out because
they would imply strong quantum geometry corrections. (In a pure
minisuperspace setting, models with suppressed quantum geometry
corrections can be constructed consistently and rather
straightforwardly. But they do not consistently model the full
theory.) In particular, critical densities must be sufficiently
smaller than Planckian, which enlarges holonomy corrections; this may
be of interest for potential observations in cosmology because
effects, such as those in the tensor mode spectrum
\cite{TensorHalf,SuperInflTensor,BounceTensor,CHRev,TensorHalfII},
would be enlarged. This would have no effect on existing analyses of
inverse volume corrections for perturbative inhomogeneities
\cite{InhomEvolve,ConstraintAlgebra,ScalarGaugeInv} where consistent
versions were already used. Since it has become quite fashionable to
boldly extend tree-level equations such as (\ref{Tree}) and its
analogs in black hole models well beyond their proven range of
validity and deeply into putative quantum regimes around classical
singularities, one should note that this is not done here. To rule out
a Planckian critical density we do not have to consider the dynamics
near a classical singularity, nor do we have to consider long
stretches of evolution which would both give rise to quantum
back-reaction in non-solvable models. For a Planckian critical
density, consistent inverse volume corrections would be significant at
all times even if matter densities are small. In this way, reliable
upper bounds on the critical density arise.

\section*{Acknowledgements}

This work was supported in part by NSF grant 0748336.

%\bibliographystyle{preprint}
%\bibliography{Bib/QuantGra}

\begin{thebibliography}{10}

\bibitem{Rov}
C.\ Rovelli,
\newblock {\em Quantum Gravity},
\newblock Cambridge University Press, Cambridge, UK, 2004

\bibitem{ALRev}
A.\ Ashtekar and J.\ Lewandowski,
\newblock Background independent quantum gravity: A status report,
\newblock {\em Class.\ Quantum Grav.} 21 (2004) R53--R152, [gr-qc/0404018]

\bibitem{ThomasRev}
T.\ Thiemann,
\newblock {\em Introduction to Modern Canonical Quantum General Relativity},
\newblock Cambridge University Press, Cambridge, UK, 2007, [gr-qc/0110034]

\bibitem{LivRev}
M.\ Bojowald,
\newblock Loop Quantum Cosmology,
\newblock {\em Living Rev.\ Relativity} 11 (2008) 4, [gr-qc/0601085],
\newblock {\tt http://www.livingreviews.org/lrr-2008-4}

\bibitem{Weiss}
N.\ Weiss,
\newblock Constraints on Hamiltonian lattice formulations of field theories in
  an expanding universe,
\newblock {\em Phys.\ Rev.\ D} 32 (1985) 3228--3232

\bibitem{UnruhTime} W.\ Unruh, \newblock Time, gravity, and quantum
mechanics, In: {\em Time's arrows today}, Ed.: S.~F.~Savitt, pages
23--94, [gr-qc/9312027]

\bibitem{River}
T.\ Jacobson,
\newblock Trans-Planckian redshifts and the substance of the space-time river,
  [hep-th/0001085]

\bibitem{EvolvingHilbert}
R.\ Dold\'an, R.\ Gambini, and P.\ Mora,
\newblock Quantum mechanics for totally constrained dynamical systems and
  evolving Hilbert spaces,
\newblock {\em Int.\ J.\ Theor.\ Phys.} 35 (1996) 2057, [hep-th/9404169]

\bibitem{CosConst}
M.\ Bojowald,
\newblock The dark side of a patchwork universe,
\newblock {\em Gen.\ Rel.\ Grav.} 40 (2008) 639--660, [arXiv:0705.4398]

\bibitem{EffAc}
M.\ Bojowald and A.\ Skirzewski,
\newblock Effective Equations of Motion for Quantum Systems,
\newblock {\em Rev.\ Math.\ Phys.} 18 (2006) 713--745, [math-ph/0511043]

\bibitem{EffCons}
M.\ Bojowald, B.\ Sandh\"ofer, A.\ Skirzewski, and A.\ Tsobanjan,
\newblock Effective constraints for quantum systems,
\newblock {\em Rev.\ Math.\ Phys.}, to appear, [arXiv:0804.3365]

\bibitem{LoopRep}
C.\ Rovelli and L.\ Smolin,
\newblock Loop Space Representation of Quantum General Relativity,
\newblock {\em Nucl.\ Phys.\ B} 331 (1990) 80--152

\bibitem{QSDI}
T.\ Thiemann,
\newblock Quantum Spin Dynamics {(QSD)},
\newblock {\em Class.\ Quantum Grav.} 15 (1998) 839--873, [gr-qc/9606089]

\bibitem{QSDV}
T.\ Thiemann,
\newblock {QSD V}: Quantum Gravity as the Natural Regulator of Matter Quantum
  Field Theories,
\newblock {\em Class.\ Quantum Grav.} 15 (1998) 1281--1314, [gr-qc/9705019]

\bibitem{InhomLattice}
M.\ Bojowald,
\newblock Loop quantum cosmology and inhomogeneities,
\newblock {\em Gen.\ Rel.\ Grav.} 38 (2006) 1771--1795, [gr-qc/0609034]

\bibitem{RS:Ham}
C.\ Rovelli and L.\ Smolin,
\newblock The physical Hamiltonian in nonperturbative quantum gravity,
\newblock {\em Phys.\ Rev.\ Lett.} 72 (1994) 446--449, [gr-qc/9308002]

\bibitem{AQGI}
K.\ Giesel and T.\ Thiemann,
\newblock Algebraic Quantum Gravity (AQG) I. Conceptual Setup,
\newblock {\em Class.\ Quantum Grav.} 24 (2007) 2465--2497, [gr-qc/0607099]

\bibitem{CosmoWithoutInfl}
P.\ Peter and N.\ Pinto-Neto,
\newblock Cosmology without inflation,
\newblock {\em Phys.\ Rev.\ D} 78 (2008) 063506, [arXiv:0809.2022]

\bibitem{IsoCosmo}
M.\ Bojowald,
\newblock Isotropic Loop Quantum Cosmology,
\newblock {\em Class.\ Quantum Grav.} 19 (2002) 2717--2741, [gr-qc/0202077]

\bibitem{AshVarReell}
J.~F.\ Barbero~G.,
\newblock Real Ashtekar Variables for Lorentzian Signature Space-Times,
\newblock {\em Phys.\ Rev.\ D} 51 (1995) 5507--5510, [gr-qc/9410014]

\bibitem{Immirzi}
G.\ Immirzi,
\newblock Real and Complex Connections for Canonical Gravity,
\newblock {\em Class.\ Quantum Grav.} 14 (1997) L177--L181

\bibitem{Bohr}
A.\ Ashtekar, M.\ Bojowald, and J.\ Lewandowski,
\newblock Mathematical structure of loop quantum cosmology,
\newblock {\em Adv.\ Theor.\ Math.\ Phys.} 7 (2003) 233--268, [gr-qc/0304074]

\bibitem{LQCStepping}
C.\ Rovelli and F.\ Vidotto,
\newblock Stepping out of Homogeneity in Loop Quantum Cosmology,
\newblock {\em Class.\ Quantum Grav.} 25 (2008) 225024, [arXiv:0805.4585]

\bibitem{APSII}
A.\ Ashtekar, T.\ Pawlowski, and P.\ Singh,
\newblock Quantum Nature of the Big Bang: Improved dynamics,
\newblock {\em Phys.\ Rev.\ D} 74 (2006) 084003, [gr-qc/0607039]

\bibitem{BounceArea}
P.\ Dzierzak, J.\ Jezierski, P.\ Malkiewicz, and W.\ Piechocki,
\newblock Quantum Big Bounce, [arXiv:0810.3172]

\bibitem{AreaVol}
C.\ Rovelli and L.\ Smolin,
\newblock Discreteness of Area and Volume in Quantum Gravity,
\newblock {\em Nucl.\ Phys.\ B} 442 (1995) 593--619, [gr-qc/9411005],
\newblock Erratum: {\em Nucl.\ Phys.\ B} 456 (1995) 753

\bibitem{Vol2}
A.\ Ashtekar and J.\ Lewandowski,
\newblock Quantum Theory of Geometry II: Volume Operators,
\newblock {\em Adv.\ Theor.\ Math.\ Phys.} 1 (1997) 388--429, [gr-qc/9711031]

\bibitem{InvScale}
M.\ Bojowald,
\newblock Inverse Scale Factor in Isotropic Quantum Geometry,
\newblock {\em Phys.\ Rev.\ D} 64 (2001) 084018, [gr-qc/0105067]

\bibitem{VolNum}
J.\ Brunnemann and T.\ Thiemann,
\newblock Simplification of the Spectral Analysis of the Volume Operator in
  Loop Quantum Gravity,
\newblock {\em Class.\ Quantum Grav.} 23 (2006) 1289--1346, [gr-qc/0405060]

\bibitem{BoundFull}
J.\ Brunnemann and T.\ Thiemann,
\newblock Unboundedness of Triad-Like Operators in Loop Quantum Gravity,
\newblock {\em Class.\ Quantum Grav.} 23 (2006) 1429--1483, [gr-qc/0505033]

\bibitem{VolSpecI}
J.\ Brunnemann and D.\ Rideout,
\newblock Properties of the Volume Operator in Loop Quantum Gravity I: Results,
\newblock {\em Class.\ Quant.\ Grav.} 25 (2008) 065001, [arXiv:0706.0469]

\bibitem{VolSpecII}
J.\ Brunnemann and D.\ Rideout,
\newblock Properties of the Volume Operator in Loop Quantum Gravity II:
  Detailed Presentation,
\newblock {\em Class.\ Quant.\ Grav.} 25 (2008) 065002, [arXiv:0706.0382]

\bibitem{SchwarzN}
M.\ Bojowald, D.\ Cartin, and G.\ Khanna,
\newblock Lattice refining loop quantum cosmology, anisotropic models and
  stability,
\newblock {\em Phys.\ Rev.\ D} 76 (2007) 064018, [arXiv:0704.1137]

\bibitem{Ambig}
M.\ Bojowald,
\newblock Quantization ambiguities in isotropic quantum geometry,
\newblock {\em Class.\ Quantum Grav.} 19 (2002) 5113--5130, [gr-qc/0206053]

\bibitem{ICGC}
M.\ Bojowald,
\newblock Loop Quantum Cosmology: Recent Progress, {\em Pramana} 63 (2004)
765--776,
\newblock In {\em Proceedings of the International Conference on Gravitation
  and Cosmology (ICGC 2004), Cochin, India}, [gr-qc/0402053]

\bibitem{QuantCorrPert}
M.\ Bojowald, H.\ Hern\'andez, M.\ Kagan, and A.\ Skirzewski,
\newblock Effective constraints of loop quantum gravity,
\newblock {\em Phys.\ Rev.\ D} 75 (2007) 064022, [gr-qc/0611112]

\bibitem{ConstraintAlgebra}
M.\ Bojowald, G.\ Hossain, M.\ Kagan, and S.\ Shankaranarayanan,
\newblock Anomaly freedom in perturbative loop quantum gravity,
\newblock {\em Phys.\ Rev.\ D} 78 (2008) 063547, [arXiv:0806.3929]

\bibitem{BouncePot}
M.\ Bojowald, H.\ Hern\'andez, and A.\ Skirzewski,
\newblock Effective equations for isotropic quantum cosmology including matter,
\newblock {\em Phys.\ Rev.\ D} 76 (2007) 063511, [arXiv:0706.1057]

\bibitem{BounceSqueezed}
M.\ Bojowald,
\newblock How quantum is the big bang?,
\newblock {\em Phys.\ Rev.\ Lett.} 100 (2008) 221301, [arXiv:0805.1192]

\bibitem{RSLoopDual}
P.\ Singh,
\newblock Loop cosmological dynamics and dualities with Randall-Sundrum
  braneworlds,
\newblock {\em Phys.\ Rev.\ D} 73 (2006) 063508, [gr-qc/0603043]

\bibitem{QuantumBounce}
M.\ Bojowald,
\newblock Quantum nature of cosmological bounces,
\newblock {\em Gen.\ Rel.\ Grav.} (2008) to appear, [arXiv:0801.4001]

\bibitem{BouncePert}
M.\ Bojowald,
\newblock Large scale effective theory for cosmological bounces,
\newblock {\em Phys.\ Rev.\ D} 75 (2007) 081301(R), [gr-qc/0608100]

\bibitem{AmbigBounce}
O.\ Hrycyna, J.\ Mielczarek, and M.\ Szyd{\l}owski,
\newblock Effects of the quantisation ambiguities on the Big Bounce dynamics,
  [arXiv:0804.2778]

\bibitem{Gamma}
M.\ Domagala and J.\ Lewandowski,
\newblock Black hole entropy from Quantum Geometry,
\newblock {\em Class.\ Quantum Grav.} 21 (2004) 5233--5243, [gr-qc/0407051]

\bibitem{Gamma2}
K.~A.\ Meissner,
\newblock Black hole entropy in Loop Quantum Gravity,
\newblock {\em Class.\ Quantum Grav.} 21 (2004) 5245--5251, [gr-qc/0407052]

\bibitem{BounceCohStates}
M.\ Bojowald,
\newblock Dynamical coherent states and physical solutions of quantum
  cosmological bounces,
\newblock {\em Phys.\ Rev.\ D} 75 (2007) 123512, [gr-qc/0703144]

\bibitem{BeforeBB}
M.\ Bojowald,
\newblock What happened before the big bang?,
\newblock {\em Nature Physics} 3 (2007) 523--525

\bibitem{Harmonic}
M.\ Bojowald,
\newblock Harmonic cosmology: How much can we know about a universe before the
  big bang?,
\newblock {\em Proc.\ Roy.\ Soc.\ A} 464 (2008) 2135--2150, [arXiv:0710.4919]

\bibitem{RefinementInflation}
W.\ Nelson and M.\ Sakellariadou,
\newblock Lattice Refining Loop Quantum Cosmology and Inflation,
\newblock {\em Phys.\ Rev.\ D} 76 (2007) 044015, [arXiv:0706.0179]

\bibitem{RefinementMatter}
W.\ Nelson and M.\ Sakellariadou,
\newblock Lattice Refining LQC and the Matter Hamiltonian,
\newblock {\em Phys.\ Rev.\ D} 76 (2007) 104003, [arXiv:0707.0588]

\bibitem{tensor}
M.\ Bojowald and G.\ Hossain,
\newblock Quantum gravity corrections to gravitational wave dispersion,
\newblock {\em Phys.\ Rev.\ D} 77 (2008) 023508, [arXiv:0709.2365]

\bibitem{FermionBBN}
M.\ Bojowald, R.\ Das, and R.\ Scherrer,
\newblock Dirac fields in Loop Quantum Gravity and Big Bang Nucleosynthesis,
\newblock {\em Phys.\ Rev.\ D} 77 (2008) 084003, [arXiv:0710.5734]

\bibitem{BHInt}
A.\ Ashtekar and M.\ Bojowald,
\newblock Quantum Geometry and the Schwarzschild Singularity,
\newblock {\em Class.\ Quantum Grav.} 23 (2006) 391--411, [gr-qc/0509075]

\bibitem{ModestoConn}
L.\ Modesto,
\newblock Loop quantum black hole,
\newblock {\em Class.\ Quantum Grav.} 23 (2006) 5587--5601, [gr-qc/0509078]

\bibitem{TensorHalf}
A.\ Barrau and J.\ Grain,
\newblock Holonomy corrections to the cosmological primordial tensor power
  spectrum, [arXiv:0805.0356]

\bibitem{SuperInflTensor}
E.~J.\ Copeland, D.~J.\ Mulryne, N.~J.\ Nunes, and M.\ Shaeri,
\newblock The gravitational wave background from super-inflation in Loop
  Quantum Cosmology, [arXiv:0810.0104]

\bibitem{BounceTensor}
J.\ Mielczarek,
\newblock Gravitational waves from the Big Bounce, [arXiv:0807.0712]

\bibitem{CHRev}
G.\ Calcagni and G.\ Hossain,
\newblock Loop quantum cosmology and tensor perturbations in the early
  universe,
\newblock {\em Adv.\ Sci.\ Lett.} (2008) to appear, [arXiv:0810.4330]

\bibitem{TensorHalfII}
A.\ Barrau and J.\ Grain,
\newblock Cosmological footprint of loop quantum gravity, in preparation

\bibitem{InhomEvolve}
M.\ Bojowald, H.\ Hern\'andez, M.\ Kagan, P.\ Singh, and A.\ Skirzewski,
\newblock Formation and evolution of structure in loop cosmology,
\newblock {\em Phys.\ Rev.\ Lett.} 98 (2007) 031301, [astro-ph/0611685]

\bibitem{ScalarGaugeInv}
M.\ Bojowald, G.\ Hossain, M.\ Kagan, and S.\ Shankaranarayanan,
\newblock Gauge invariant cosmological perturbation equations with corrections
  from loop quantum gravity, [arXiv:0811.1572]

\end{thebibliography}

\end{document}